\documentclass{PoS}
\usepackage{amsmath}
\usepackage{amsfonts}
\usepackage{textcomp}
\usepackage{graphicx}
\usepackage{dcolumn}
\usepackage{tikz,etoolbox}
\usepackage[point,rounding]{rccol}
\usepackage{booktabs}
\usepackage{varwidth}
\usepackage{colortbl}
\usepackage{color}
\newcommand{\pTm}{p_{\mbox{\tiny T}}}
\newcommand{\mTm}{m_{\mbox{\tiny T}}}

\newcommand{\Pionm}{\pi^{0}}

\newcommand{\Figure}[1]    {Figure~\ref{#1}}
\newcommand{\Ref}[1]       {Ref.~\cite{#1}}

\title{Direct Photon Production and Flow at Low Transverse Momenta in pp, p--Pb and Pb--Pb Collisions}

\ShortTitle{$\gamma_\text{dir}$ and flow in pp, p--Pb and Pb--Pb with ALICE}

\author{\speaker{Nicolas Schmidt}\textbf{, on behalf of the ALICE collaboration}\\
Oak Ridge National Laboratory \& IKF Frankfurt\\
E-mail: \email{nicolas.schmidt@cern.ch}}

\abstract{Low transverse momentum direct photon measurements have been carried out by the ALICE experiment at the CERN LHC in small collision systems (pp, $\sqrt{s}=2.76$ and 8 TeV and p--Pb, $\sqrt{s_\text{NN}}=5.02$ TeV) as well as in heavy-ion collisions (Pb--Pb, $\sqrt{s_\text{NN}}=2.76$ TeV). For the first time, also the multiplicity dependence of direct photon production was investigated in p--Pb collisions. Whereas in the small systems no significant thermal photon signal was observed, a $\sim15\%$ excess has been measured in central Pb--Pb collisions in the region of $\pTm<3$ GeV/$c$. A signal of prompt photon production at high transverse momentum consistent with binary scaling has been observed in all collision systems following NLO pQCD predictions. Direct photon flow has been measured in central and semi-central Pb--Pb collisions and found to be of similar size as the charged hadron and decay photon flow.}

\FullConference{International Conference on Hard and Electromagnetic Probes of High-Energy Nuclear Collisions\\
30 September - 5 October 2018\\
Aix-Les-Bains, Savoie, France}

\begin{document}

  \section{Introduction}
  In high-energy hadron collisions, direct photons are created during all stages and can leave the medium unaffected due to their negligible final state interactions making them an ideal probe to extract information about this medium. They are categorized depending on their production mechanisms. In all collision systems a signal of prompt photons is expected. These photons are produced at leading order by Compton scattering, quark annihilation and parton fragmentation which are well described by NLO pQCD calculations. Measurements of prompt photons can be used to test binary scaling in p--Pb or heavy-ion collisions at high $\pTm$. 
  Additional sources of direct photons are thermal photons, photons from jet-medium interactions as well as pre-equilibrium photons. Thermal photons are produced via scattering of thermalized particles and are dominant at low $\pTm$ following an exponentially decreasing spectral shape. For thermal photons, depending on their production, also a large collective flow ($v_2$) is expected. Measurements of the yield and flow of thermal photons therefore provide information about the properties and the space-time evolution of the medium produced in heavy-ion collisions.
  
  \section{Detector description and datasets}
  Three independent reconstruction techniques are used to measure inclusive photons in the central barrel of ALICE \cite{Abelev:2014ffa} consisting of two calorimetric methods using the EMCal or the PHOS detector and the photon conversion method (PCM). The latter reconstructs photons that have converted within the detector material from the resulting e$^+$e$^-$ pairs using the Inner Tracking System (ITS) and the Time Projection Chamber (TPC). The presented measurements use pp ($\sqrt{s}=2.76$ and 8 TeV), p--Pb ($\sqrt{s_\text{NN}}=5.02$ TeV) and Pb--Pb ($\sqrt{s_\text{NN}}=2.76$ TeV) collision data recorded in 2010-2013. The following detector descriptions cover their configurations during this time. 
  
  The EMCal is a lead-scintillator sampling calorimeter with an acceptance of $\Delta\varphi=100^\circ$ and $|\eta|<0.7$. Each of its 11,520 cells has a size of $\Delta\eta\times\Delta\varphi=0.0143\times0.0143$ rad$^2$ providing an energy resolution of $\sigma_E/E=4.8\%/E\oplus11.3\%/\sqrt{E}\oplus1.7\%$ with $E$ in units of GeV.
  The PHOS is a lead tungstate (PbWO$_4$) crystal calorimeter covering $\Delta\varphi=60^\circ$ and $|\eta|<0.12$. Its cells have a size of $22\times22$ mm$^2$ which is slightly larger than the Moli\`{e}re radius of 2 cm. PHOS achieves an energy resolution of $\sigma_E/E=1.8\%/E\oplus3.3\%/\sqrt{E}\oplus1.1\%$.
  The ITS is a six layered silicon detector system made of Silicon Pixel Detectors (SPD), Silicon Drift Detectors (SDD) and Silicon Strip Detectors (SSD). It is used for charged particle track reconstruction, particle identification and primary vertex reconstruction. The ITS covers the full azimuthal angle and has an acceptance in pseudorapidity of $|\eta|<2$ and $|\eta|<1.4$ for the two SPD layers and $|\eta|<0.7$ for the SDD and SSD layers.
  The TPC is a drift chamber covering the full azimuth and $|\eta|<0.9$. It is filled with a gas mixture of Ne-CO$_2$ (90\%-10\%) and allows tracking and particle identification of charged particles down to $\pTm\approx100$ MeV/$c$ and $50$ MeV/$c$ for primaries or secondaries, respectively. 
  The V0 detector is used as a trigger for minimum bias events and centrality estimation in p--Pb and Pb--Pb collisions. It consists of two scintillator arrays (V0A and V0C) covering $2.8<\eta<5.1$ and $-3.7<\eta<-1.7$, respectively. The minimum bias trigger (MB$_\text{OR}$) condition required a hit in either the SPD, V0A or the V0C or a coincidence between V0A and V0C (MB$_\text{AND}$) depending on the year.
  
  \section{Photon reconstruction in ALICE}
  \label{sec:photonreco}
  The PCM method uses the tracking of charged particles and subsequent secondary vertex finding followed by particle identification and finally photon candidate reconstruction and selection. Its efficiency depends on the conversion probability in the inner detector material of ALICE of about 8.9\%. The analysis selection criteria are described in detail for the pp and Pb--Pb measurements in the respective publications \cite{Acharya:2018dqe,Adam:2015lda,Acharya:2018bdy}. The p--Pb measurements follow the same criteria as the pp measurement at $\sqrt{s}=8$ TeV.
  Photon reconstruction with the calorimeters PHOS and EMCal is based on the measured energy deposit from the electromagnetic shower that is created when a photon or electron/positron enters the calorimeter. The shower usually spreads to adjacent cells which requires a clustering of cells to reconstruct the full shower energy.
  
  \section{Direct photon and flow measurements}

  As direct photons can not be experimentally distinguished from decay photons, the presented signal extraction method follows a subtraction method. For this, the inclusive photons are measured and the decay photons are subtracted. The decay photons are obtained from a particle decay simulation that uses parameterizations of measured or $\mTm$-scaled hadron spectra in the respective collision system as input. 
  In order to quantify the direct photon signal and to have a better handle on systematic uncertainties the direct photon excess ratio $R_{\gamma} = \frac{\gamma_{\text{inc}}}{\gamma_{\text{dec}}} \approx \left(\frac{\gamma_{\text{inc}}}{\pi^{0}}\right)_\text{meas} / \left(\frac{\gamma_{\text{decay}}}{\pi^{0}}\right)_\text{sim}$ is used.
 
 \begin{figure}
    \center
    \includegraphics[height=.24\textheight]{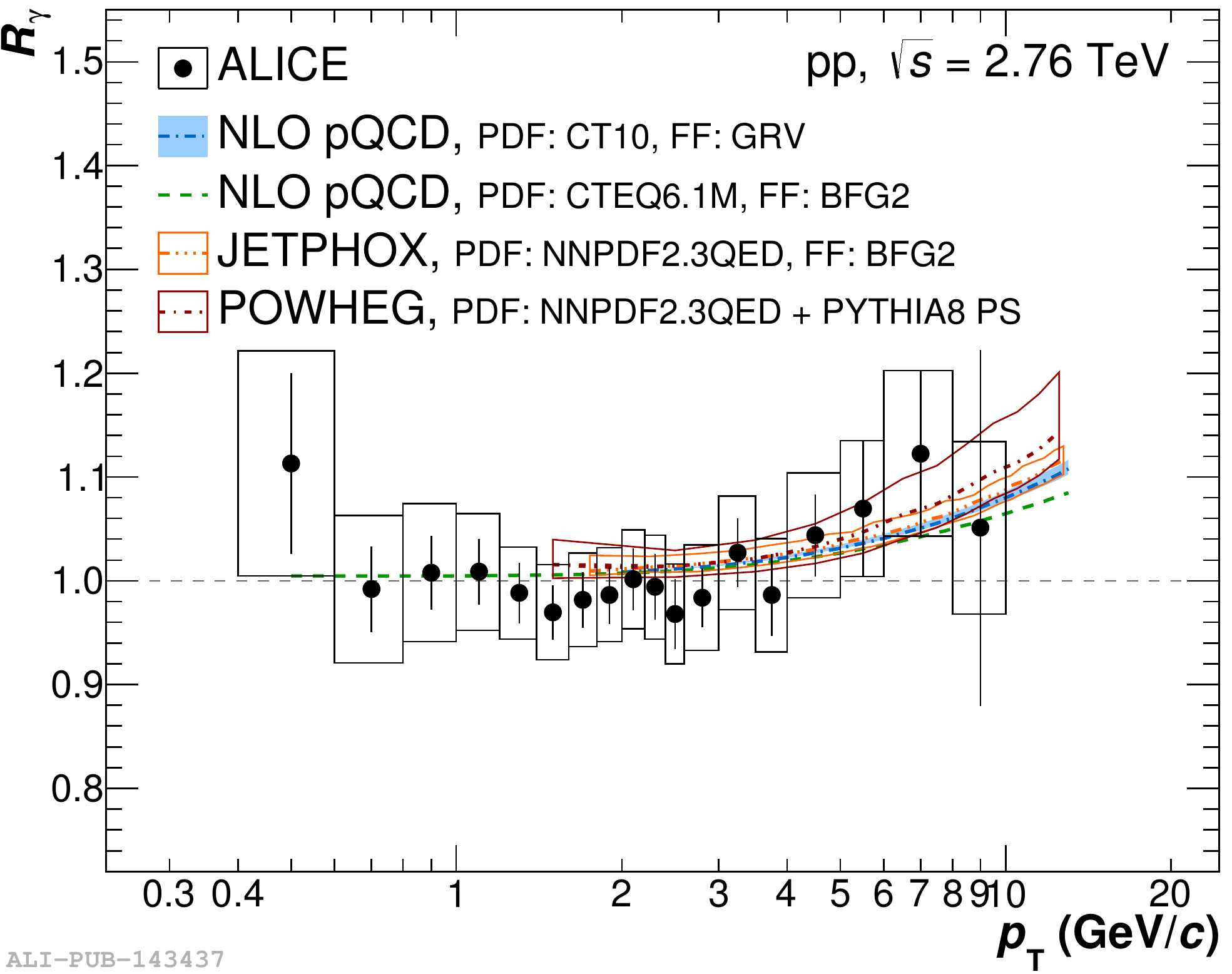}
    \hspace{0.7cm}
    \includegraphics[height=.24\textheight]{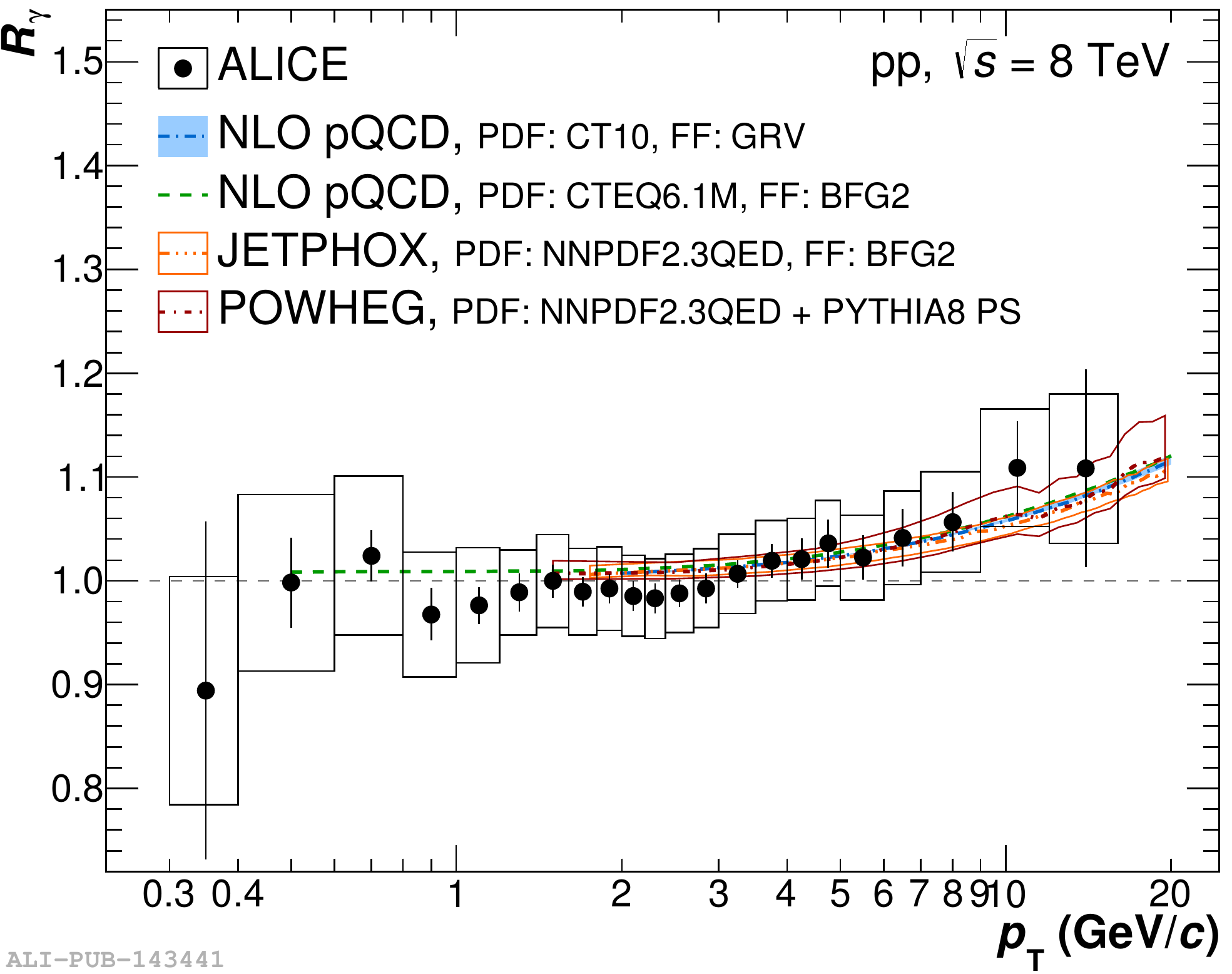}
      \caption{Direct photon excess ratios for the combined measurements from pp collisions at $\sqrt{s}=2.76$ TeV (Left) and 8 TeV (right) together with NLO pQCD calculations \cite{Paquet:2015lta,Jager:2002xm} as well as JETPHOX and POWHEG \cite{Klasen:2017dsy} predictions.}
    \label{fig:ppdirgamma}
  \end{figure}
 
  Inclusive and direct photon measurements have been carried out in pp ($\sqrt{s}=2.76$ and 8 TeV \cite{Acharya:2018dqe}), p--Pb ($\sqrt{s_\text{NN}}=5.02$ TeV) and Pb--Pb ($\sqrt{s_\text{NN}}=2.76$ TeV \cite{Adam:2015lda}) collisions.
  Each analysis was performed with several partially independent reconstruction techniques, whose results were combined using the ``Best Linear Unbiased Estimates'' (BLUE) method where possible statistical and systematic uncertainty correlations are treated consistently.

    \begin{figure}
    \center
    \includegraphics[height=.37\textheight]{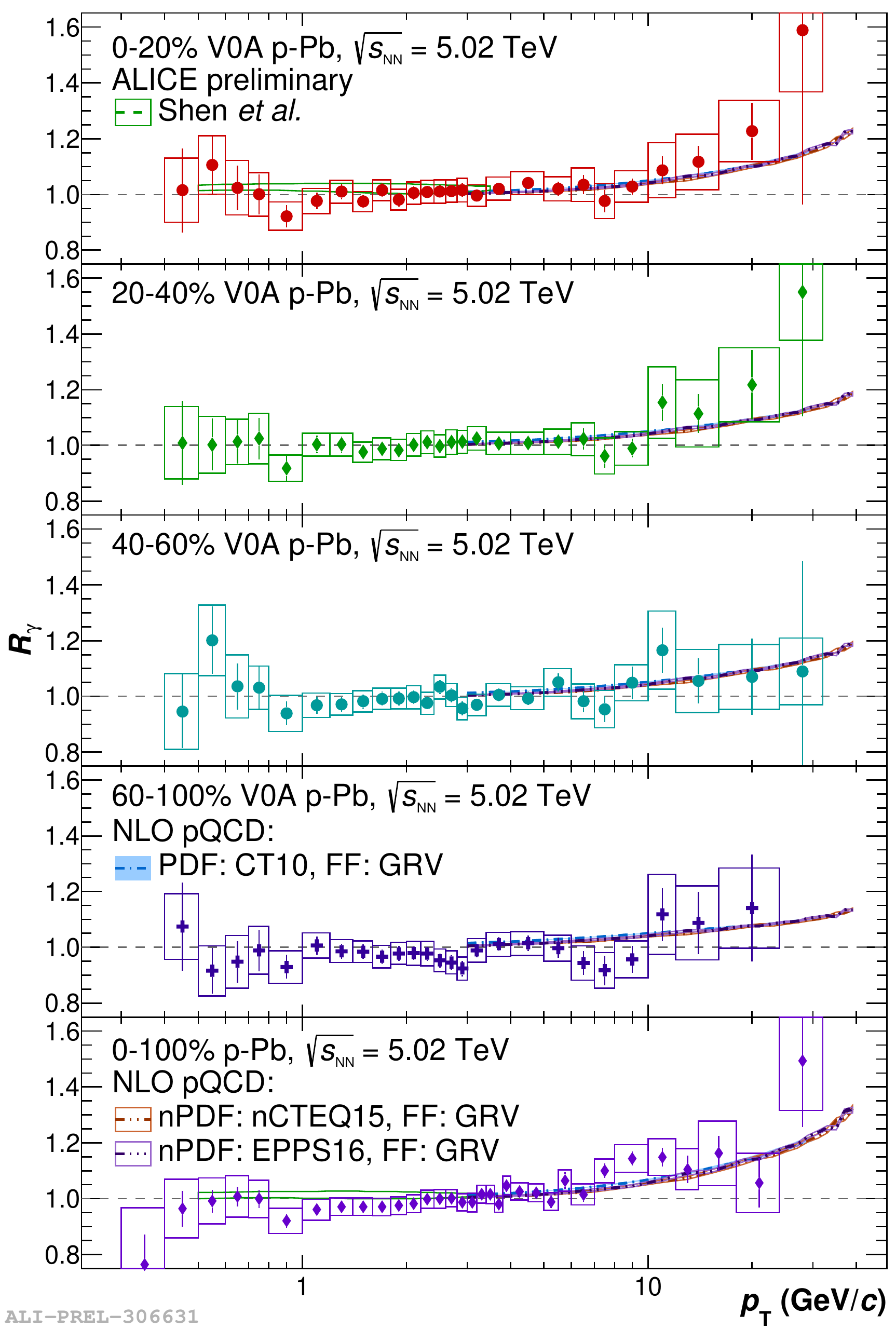}
    \hspace{0.2cm}
    \includegraphics[height=.37\textheight]{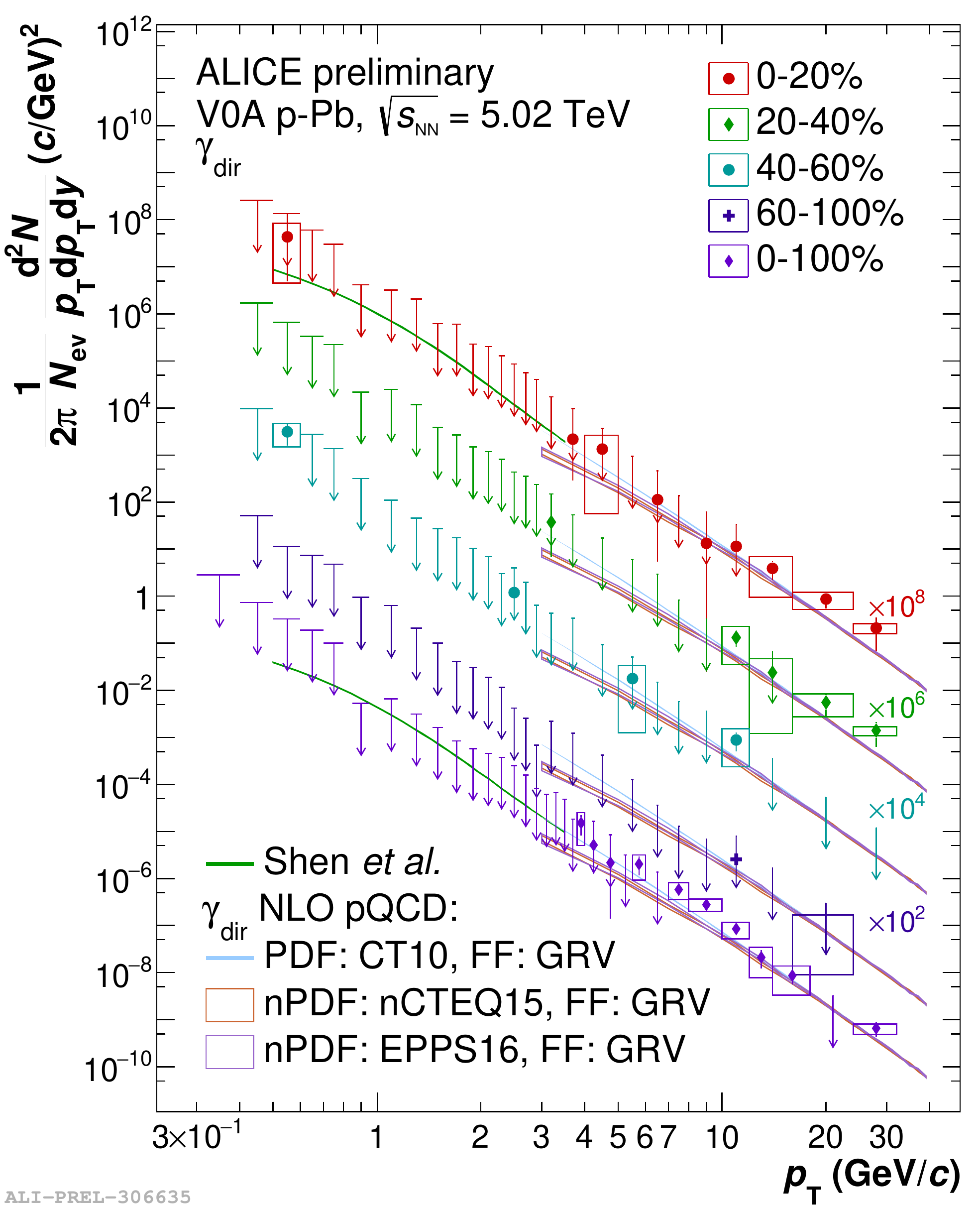}
      \caption{Left: Direct photon excess ratios for the combined measurements from p--Pb collisions at $\sqrt{s_\text{NN}}=5.02$ TeV in four multiplicity classes as well as the full NSD sample together with NLO pQCD calculations \cite{Paquet:2015lta,Jager:2002xm} as well as JETPHOX and POWHEG \cite{Klasen:2017dsy} predictions.. Right: Direct photon yield in four multiplicity classes as well as the full NSD sample. The end bars on the arrows indicate upper limits at 90\% C.L. calculated for all points where the total uncertainty of $R_\gamma$ was found to be consistent with unity.}
    \label{fig:pPbCent}
  \end{figure}
  
\begin{figure}
    \center
    \includegraphics[height=.27\textheight]{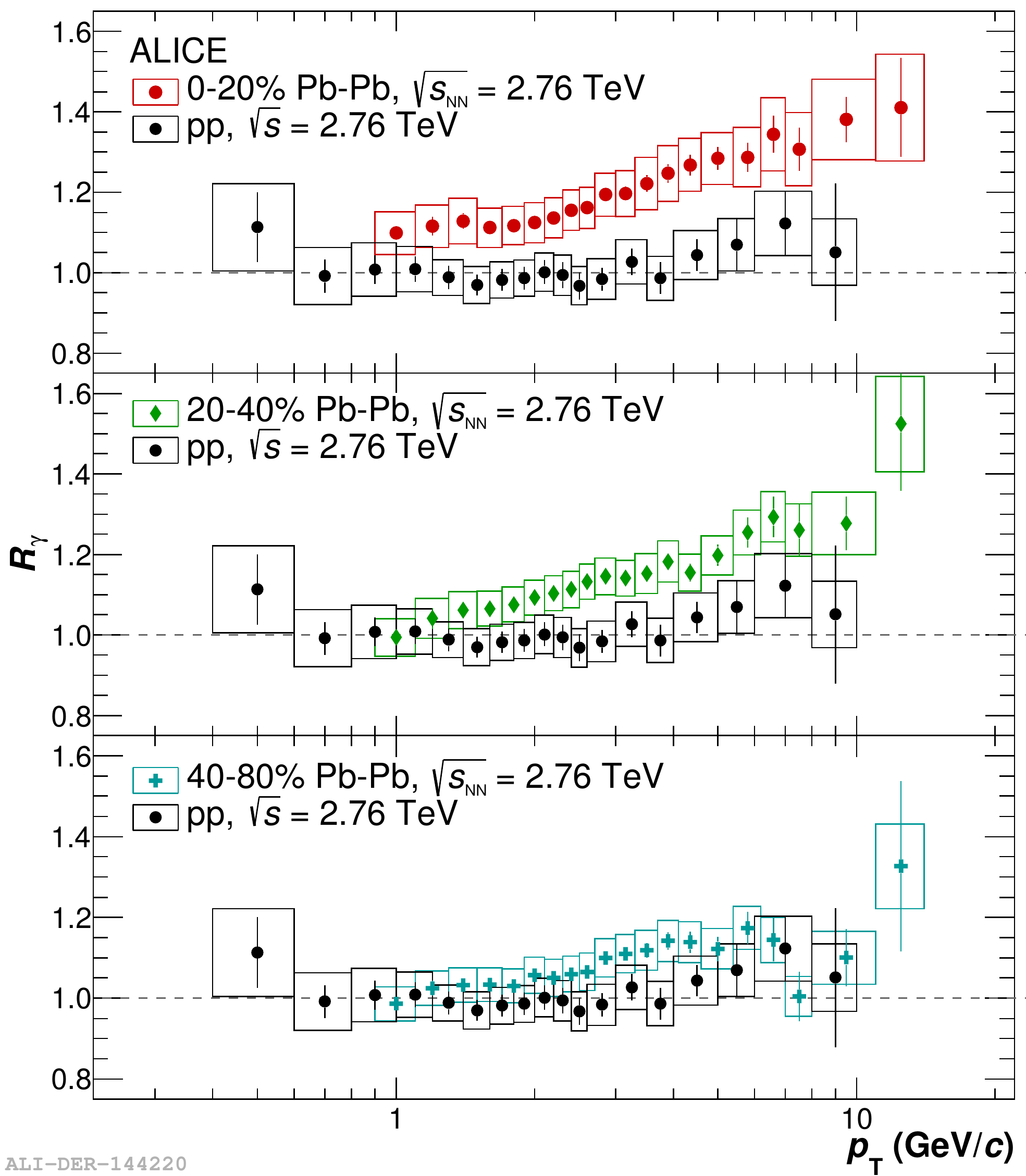}
    \hspace{0.4cm}
    \includegraphics[height=.27\textheight]{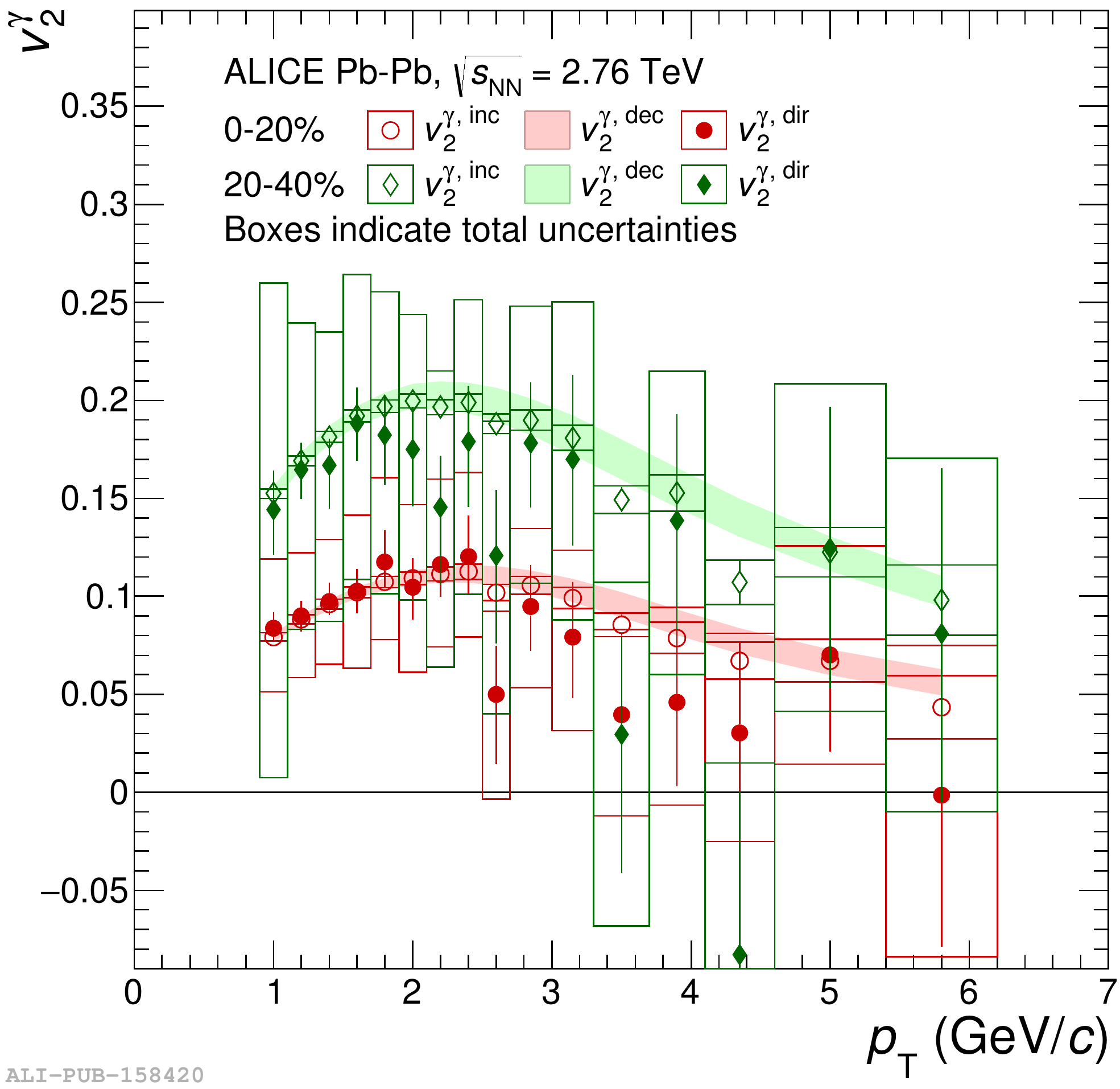}
      \caption{Left: Direct photon excess ratios for the combined measurements from Pb--Pb collisions at $\sqrt{s_\text{NN}}=2.76$ TeV in three centrality classes overlayed with the pp measurement at $\sqrt{s}=2.76$ TeV. Right: Decay, inclusive and direct photon $v_2$ for two centrality classes in Pb--Pb collisions at $\sqrt{s_\text{NN}}=2.76$ TeV. Vertical bars indicate statistical uncertainties and boxes systematic uncertainties.}
    \label{fig:PbPbandFlow}
  \end{figure}

  The direct photon excess ratios in pp collisions at $\sqrt{s}=2.76$ and 8 TeV were obtained from a combination of the PCM, EMCal and the hybrid method PCM-EMCal measurements and are shown in \Figure{fig:ppdirgamma}. For p--Pb collisions at $\sqrt{s_\text{NN}}=5.02$ TeV the PHOS-based measurement was also included and the excess ratio as well as the direct photon yield or upper limits at 90\% confidence level are given in four V0A multiplicity classes as well as the full sample of non single-diffractive (NSD) p--Pb collisions in \Figure{fig:pPbCent}. In all small systems, pp and p--Pb, no significant thermal photon signal is found within the uncertainties at $\pTm<5$ GeV/$c$. For $\pTm>6$ GeV/$c$, the onset of prompt photon production consistent with NLO pQCD calculations is observed.
  The Pb--Pb measurement at $\sqrt{s_\text{NN}}=2.76$ TeV, given in \Figure{fig:PbPbandFlow} (left) shows a thermal photon excess of $\sim15\%$ in central and $\sim9\%$ in semi-central collisions as well as a strong prompt photon signal consistent with theory predictions at high $\pTm$. The comparison to pp collisions at the same center of mass energy shows that the medium effects in heavy-ion collisions result in an enhanced signal.
  
  Flow measurements in central and semi-central Pb--Pb collisions \cite{Acharya:2018bdy} were made using a decay photon $v_2$ obtained from cocktail calculations where $v_2^{\Pionm}\approx v_2^{\pi^\pm}$ is assumed and the $v_2$ of various mesons is calculated via $KE_\text{T}$ scaling from $v_2^{K^\pm}$ \cite{Abelev:2014pua}. The inclusive photon $v_2$ was measured with PCM and PHOS and corrected for background flow from impurities before being combined. The direct photon flow was calculated via $v_2^{\gamma\text{,dir}} = (R_\gamma \cdot v_2^{\gamma\text{,inc}}-v_2^{\gamma\text{,dec} } )/(R_\gamma-1)$. As the $R_\gamma$ measurements have a significance of less $2\sigma_\text{sys}$ from unity in several $\pTm$ bins, the central values and uncertainties have been calculated using Monte Carlo simulations following a Bayesian approach as described in \Ref{Acharya:2018bdy}. As seen in \Figure{fig:PbPbandFlow} (right) a large direct photon $v_2$ for $\pTm<3$ GeV/$c$ was measured with a magnitude comparable to that of the hadrons. This result may point to late production times of the direct photons after the flow is already established or to an additional source of photons.

  \section{Conclusions}
  Direct photon measurements have been carried out in small systems as well as in heavy-ion collisions in ALICE. Several partially independent reconstruction techniques were used and combined to cover large transverse momentum regions with reduced uncertainties. For the first time, also a multiplicity dependent direct photon measurement was made in p--Pb collisions at $\sqrt{s_\text{NN}}=5.02$ TeV in four centrality classes. In all small systems no significant direct photon excess was observed in the thermal photon region of $\pTm<3$ GeV/$c$. An onset of prompt photon production with a significance of $1-2\sigma$ consistent with NLO pQCD calculations was found in all systems towards high $\pTm$.
  In Pb--Pb collisions at $\sqrt{s_\text{NN}}=2.76$ TeV a thermal photon excess of $\sim15\%$ in central and $\sim9\%$ in semi-central collisions was measured and found to be consistent with theory predictions. The direct photon $v_2$ was determined and found to be of similar size as the charged hadron flow and the inclusive photon flow. The flow measurements have a significance of 1.4$\sigma$ or 1.0$\sigma$ in the range $0.9<\pTm<2.1$ GeV/c for $0-20\%$ and $20-40\%$, respectively.

\end{document}